\documentclass[english,12pt]{article}
\usepackage[a4paper]{geometry}
\usepackage[T1]{fontenc}
\usepackage[latin9]{inputenc}
\usepackage{amsthm}
\usepackage{amsmath}
\usepackage{amsfonts}
\usepackage{babel}

\usepackage{mathptmx}
\usepackage{helvet}
\usepackage{courier}  
\usepackage{eurosym} 
\usepackage{type1cm}
\usepackage{url}

\usepackage{makeidx}         
\usepackage{graphicx}        
\usepackage{multicol}        
\usepackage[bottom]{footmisc}

\usepackage{amssymb,amsmath,amstext}
\usepackage{comment}
\usepackage{rotating}
\usepackage[all]{xy}
\usepackage{enumerate}

\usepackage{pifont}

                          {\end{enumerate}}

\newtheorem{definition}{Definition}

\begin{document}

\title{Dissipativity in economic model predictive control: beyond steady-state optimality}

\author{Matthias A. M\"uller\\[.2cm]\normalsize Institute of Automatic Control, Leibniz University Hannover, 30167 Hannover, Germany \\ \normalsize mueller@irt.uni-hannover.de}

\date{}
\maketitle
\thispagestyle{empty}

\section{Introduction} \label{subsec_introduction}

Within the last decade, the study of economic MPC schemes has received a significant amount of attention. 
Here, in contrast to standard stabilizing or tracking MPC, some general performance criterion is considered, which is not necessarily related to the tracking of a given setpoint or (dynamic) trajectory. 
As a result, the employed cost function is not necessarily positive definite with respect to such a setpoint or trajectory.
Such general control objectives occur in many applications and can, e.g., correspond to the maximization of a certain product in the process industry (or profit maximization in general), the minimization of energy consumption, or a cost efficient scheduling of a production process, to name but a few. 
These type of applications also served as motivation for the term \emph{economic} MPC.

When considering such general cost functions in the context of MPC, various questions arise.
The first immediate issue is that of determining what the optimal operating conditions are for given system dynamics, cost function, and constraints. 
This means to assess whether it is optimal (in terms of the employed cost function) to operate the system at the best steady state, at some periodic orbit, or under some more complex operating conditions (e.g., some general set or time-varying trajectory).
The second question which directly follows is whether the closed-loop system resulting from application of an economic MPC scheme converges to the optimal regime of operation, i.e., whether the closed-loop system "'does the right thing"'. 
For example, in case of optimal periodic operation, the closed-loop system should converge to the optimal periodic orbit.

In order to answer the above questions, certain dissipativity conditions have turned out to play a crucial role in the context of economic MPC.
In the literature, first the most basic case where steady-state operation is optimal has been studied, which is by now fairly well understood.
On the other hand, results for more general cases which go beyond steady-state optimality have only been obtained recently and the picture is (at least partially) still much less complete here.

The goal of this chapter is to give a concise survey on the dissipativity conditions that have appeared in the economic MPC literature and to discuss their decisive role in this context.
The structure of this chapter is as follows.
After the presentation of the setup and a brief introduction to the concept of dissipativity (Sections~~\ref{subsec_setup} and~\ref{subsec_dissipativity}), we start with the basic case of optimal steady-state operation (Section~\ref{subsec_optimal_steady_state_operation}), and then move on to the cases of optimal periodic operation (Section~\ref{subsec_optimal_periodic_operation}) and more general optimal regimes of operation (Section~\ref{subsec_general_optimal_operation}). 
After that, we briefly discuss how the employed dissipativity conditions can be verified in Section~\ref{subsec_comp_storage_fct}.
In Section~\ref{subsec_time_varying}, we discuss the time-varying case, before concluding this chapter in Section~\ref{subsec_conclusions}.
We remark that we do not aim at providing a survey on all existing economic MPC approaches, but focus on the role played by dissipativity in the context of economic MPC.
For a more comprehensive introduction to economic MPC in general, the interested reader is, e.g., referred to the recent survey article~\cite{Faulwasser_Grune_Muller_18} or book~\cite{Ellis17a}.  

\section{Setup} \label{subsec_setup}
We consider discrete-time nonlinear systems\footnote{Some of the following results are also available for continuous-time systems, compare, e.g.,~\cite{Faulwasser_et_al_CDC14,Faulwasser_et_al_Aut17}.} of the form
\begin{equation}
	x(k+1) = f(x(k),u(k))\text{, }\qquad x(0) = x^0\text{,} \label{system}
\end{equation}
with $k \in \mathbb{N}_0$. 
The input and state are required to satisfy pointwise-in-time input and state constraints, respectively, i.e.\footnote{Also coupled constraints of the form $(x,u)\in\mathbb{Z}\subseteq\mathbb{R}^n\times\mathbb{R}^m$ can be considered.}, $u(k)\in\mathbb{X}$ and 
$x(k) \in \mathbb{X}$ for all $k \in \mathbb{N}_0$.
Define $\mathcal{U}_N(x):=\{\textbf{u}\in\mathbb{U}^N:x_{\textbf{u}}(k;x)\in\mathbb{X}~\textnormal{for all}~k=0,\dots,N\}$ for $N\in\mathbb{N}\cup\{\infty\}$ as the set of feasible input sequences of length~$N$.
The performance criterion to be optimized is specified by the stage cost function $\ell: \mathbb{R}^n \times \mathbb{R}^m \rightarrow \mathbb{R}$, which is assumed to be continuous. 
As already discussed in Section~\ref{subsec_introduction}, in economic MPC no conditions (other than continuity) are imposed on the stage cost $\ell$, in particular no positive definiteness requirements.
The MPC optimization problem at time $t$, given the current measured state $\hat{x}:=x(t)$, is then given by
\begin{align}
	V(\hat{x}) = 
		\inf_{\textbf{u}\in\mathcal{U}_N(\hat{x})} \sum_{k=0}^{N-1} \ell(x_{\mathbf{u}}(k;\hat{x}),u(k)),  \label{economic_MPC_problem}
\end{align}
and the input $u(t)=u^\ast(0;\hat{x})$ is applied to system~\eqref{system} in a standard receding horizon fashion.
Various existing economic MPC schemes employ additional terminal constraints and/or a terminal cost function in~\eqref{economic_MPC_problem}. 
The dissipativity conditions which are discussed in the following play a crucial role in both settings with and without such additional terminal ingredients.

\section{Dissipativity} \label{subsec_dissipativity}
The concept of dissipativity was introduced by Willems in~\cite{Willems_Dissipativity_PartI_72} (compare also \cite{Byrnes_Lin_TAC94} for a discrete-time version) and is as follows.
\begin{definition} \label{def_dissipativity}
The system~(\ref{system}) is dissipative on $\mathbb{X}\times\mathbb{U}$ with respect to the supply rate $s:\mathbb{R}^n\times\mathbb{R}^m\rightarrow\mathbb{R}$ if there exists a storage function~$\lambda:\mathbb{R}^n\rightarrow\mathbb{R}_{\geq 0}$ such that the following inequality is satisfied for all $(x,u)\in\mathbb{X}\times\mathbb{U}$ with $f(x,u)\in\mathbb{X}$:
\begin{align}
\lambda(f(x,u))-\lambda(x)\leq s(x,u).  \label{dissipation_inequality}
\end{align}
If there exists $\rho\in\mathcal{K}_{\infty}$ and a set $\mathbb{X}^*\subseteq\mathbb{X}$ such that for all $(x,u)\in\mathbb{X}\times\mathbb{U}$ with $f(x,u)\in\mathbb{X}$ it holds that\footnote{Here, $|x|_{\mathbb{X}^*}$ denotes the distance of the point~$x$ to the set~$\mathbb{X}^*$, i.e., $|x|_{\mathbb{X}^*}:=\min_{y\in\mathbb{X}^*}|x-y|.$}
\begin{align}
\lambda(f(x,u))-\lambda(x)\leq -\rho(|x|_{\mathbb{X}^*})+s(x,u),  \label{strict_dissipation_inequality}
\end{align}
then system~(\ref{system}) is strictly dissipative with respect to the supply rate~$s$ and the set~$\mathbb{X}^*$. 
\end{definition}
Interpreting the storage function~$\lambda$ as "`generalized energy"', the property of dissipativity means that along any solution of system~\eqref{system}, energy is dissipated, i.e., the difference in stored energy is less than or equal to what is supplied to the system from the outside (via the supply rate~$s$).
Analogously, strict dissipativity means that energy is strictly dissipated along all trajectories which do not completely lie inside the set $\mathbb{X}^*$. 

In the following sections, we discuss how dissipativity (or strict dissipativity) with respect to different supply rates (and different sets $\mathbb{X}^*$) can be employed in the context of economic MPC, both for classifying different optimal operating conditions as well as closed-loop convergence.

\section{Optimal steady-state operation} \label{subsec_optimal_steady_state_operation}
The paper~\cite{Angeli_Amrit_Rawlings_TAC12} was the first\footnote{In the earlier work~\cite{Diehl_Amrit_Rawlings_TAC11}, a special case of a dissipativity condition requiring a linear storage function has already been used, although the employed assumption was not recognized as a dissipativity condition.} paper where connections between dissipativity and economic MPC have been made apparent.
In that work, the most basic case was considered where the optimal operating regime for system~\eqref{system} is steady-state operation.
In order to formally define this notion, we first define an optimal steady-state and input pair $(x^*,u^*)$ as a minimizer\footnote{Given continuity of~$\ell$, such a minimizer exists, e.g., if $\mathbb{X}$ and $\mathbb{U}$ are compact or if $\ell$ is radially unbounded. If the minimizer is not unique, in the following $(x^*,u^*)$ denotes an arbitrary of the multiple minimizers.} to the following optimization problem:
\begin{align*}
\min_{x\in\mathbb{X},u\in\mathbb{U},x=f(x,u)}\ell(x,u)
\end{align*}
The property of optimal steady-state operation can now be defined as follows~\cite{Angeli_Amrit_Rawlings_TAC12}:
\begin{definition} \label{def_optimal_steady_state_operation}
The system~(\ref{system}) is \emph{optimally operated at steady-state}, if for each $x\in\mathbb{X}$ with $\mathcal{U}_{\infty}(x)\neq\emptyset$ and each $\mathbf{u}\in\mathcal{U}_{\infty}(x)$ the following holds:
\begin{align*}
\liminf_{T\rightarrow\infty}\frac{\sum_{k=0}^T\ell(x_{\textbf{u}}(k;x),u(k))}{T+1} \geq \ell(x^*,u^*). 
\end{align*}
\end{definition}
Definition~\ref{def_optimal_steady_state_operation} means that no other feasible input and state trajectory pair of system~\eqref{system} can result in an asymptotic average cost which is better than that of the optimal steady-state.

As was shown in~\cite{Angeli_Amrit_Rawlings_TAC12}, a sufficient condition for optimal steady-state operation is that system~\eqref{system} is dissipative on $\mathbb{X}\times\mathbb{U}$ with respect to the supply rate
\begin{align}
s(x,u) = \ell(x,u) - \ell(x^*,u^*). \label{supply_rate_steady_state}
\end{align}
This statement can be proven as follows. 
If the system is dissipative on $\mathbb{X}\times\mathbb{U}$ with respect to the supply rate~\eqref{supply_rate_steady_state}, we can sum up the corresponding dissipation inequality~\eqref{dissipation_inequality} along any feasible state and input sequence pair to obtain
\begin{align*}
\lambda(x_{\textbf{u}}(T;x))-\lambda(x) \leq \sum_{k=0}^{T-1}\ell(x_{\textbf{u}}(k;x),u(k))-\ell(x^*,u^*)
\end{align*}
for all $T\in\mathbb{N}_0$. 
Dividing this inequality by $T$, taking the $\liminf$ on both sides and exploiting the fact that the storage function~$\lambda$ is nonnegative, this results in
\begin{align*}
0 &\leq \liminf_{T\rightarrow\infty}\frac{\lambda(x_{\textbf{u}}(T;x))-\lambda(x)}{T} \leq \liminf_{T\rightarrow\infty}\frac{\sum_{k=0}^{T-1}\ell(x_{\textbf{u}}(k;x),u(k))-\ell(x^*,u^*)}{T} \\
&=- \ell(x^*,u^*) + \liminf_{T\rightarrow\infty}\frac{\sum_{k=0}^{T-1}\ell(x_{\textbf{u}}(k;x),u(k))}{T}. 
\end{align*}
But this implies that the system is optimally operated at steady-state according to Definition~\ref{def_optimal_steady_state_operation}.
Similarly, it was proven in~\cite{Angeli_Amrit_Rawlings_TAC12,Muller_et_al_NMPC15} that \emph{strict} dissipativity on $\mathbb{X}\times\mathbb{U}$ with respect to the supply rate~\eqref{supply_rate_steady_state} and the set $\mathbb{X}^*=\{x^*\}$ is a sufficient condition for a slightly stronger property than optimal steady-state operation. 
This slightly stronger notion was termed (uniform) suboptimal operation off steady-state and, loosely speaking, means that every other feasible input and state trajectory pair of system~\eqref{system} either results in an asymptotic average cost which is strictly worse than that of the optimal steady-state or enters a neighborhood of the optimal steady-state sufficiently often.

In two later publications~\cite{Muller_Angeli_Allgower_TAC13,Muller_et_al_NMPC15}, it was shown that under an additional (local) controllability condition, (strict) dissipativity with respect to the supply rate~\eqref{supply_rate_steady_state} (and the set $\mathbb{X}^*=\{x^*\}$) is also necessary for optimal steady-state operation (uniform suboptimal operation off steady-state).
This can be proven by a contradiction argument.
Namely, assuming optimal steady-state operation but the system being not dissipative with respect to the supply rate~\eqref{supply_rate_steady_state}, one can construct a specific feasible periodic trajectory which results in an asymptotic average cost which is strictly lower than $\ell(x^*,u^*)$, thus contradicting steady-state optimality.
In summary, under an additional controllability condition, dissipativity with respect to the supply rate~\eqref{supply_rate_steady_state} is an equivalent characterization of optimal steady-state operation.

As mentioned in Section~\ref{subsec_introduction}, a crucial question in economic MPC is to determine whether the resulting closed-loop system converges to the optimal operating regime, i.e., the optimal steady state $x^*$ in case of optimal steady-state operation.
To this end, it turns out that again the same strict dissipativity condition can be used.
The crucial observation for the stability analysis is to consider the \emph{rotated} cost function $L$, defined as
\begin{align}
L(x,u):= \ell(x,u) - \ell(x^*,u^*) + \lambda(x) - \lambda(f(x,u)), \label{rotated_cost}
\end{align}
and to note that strict dissipativity with respect to the supply rate~\eqref{supply_rate_steady_state} and the set $\mathbb{X}^*=\{x^*\}$ implies that $L$ is positive definite with respect to the optimal steady state~$x^*$.
In case that suitable additional terminal constraints are used, one can show that the optimal solution $u^\ast(\cdot;\hat{x})$ to problem~\eqref{economic_MPC_problem} is the same as that to problem~\eqref{economic_MPC_problem} with $\ell$ replaced by $L$. 
This was first shown using terminal equality constraints~\cite{Diehl_Amrit_Rawlings_TAC11,Angeli_Amrit_Rawlings_TAC12} and subsequently extended to a setting with a terminal region and terminal cost~\cite{Amrit_Rawlings_Angeli_AnnRevCont_11}.
Positive definiteness of $L$ then allows to use standard stability results from stabilizing MPC with positive definite cost functions, and hence to conclude asymptotic stability of the optimal steady state~$x^*$ for the resulting closed-loop system.
In case that no additional terminal constraints are employed, this equivalence between the optimal solutions of problem~\eqref{economic_MPC_problem} and the modified problem using $L$ instead of $\ell$ does no longer hold. 
Here, positive definiteness of the rotated cost~$L$ allows to establish certain \emph{turnpike} properties of problem~\eqref{economic_MPC_problem}, which in turn can be used to establish (practical) stability of the resulting closed-loop system, compare~\cite{Gruene_econ_MPC_sub_11,Grune_Stieler_14}.
In fact, it turns out that under certain conditions, strict dissipativity with respect to the supply rate~\eqref{supply_rate_steady_state} and the set $\mathbb{X}^*=\{x^*\}$ and the turnpike property of problem~\eqref{economic_MPC_problem} at the optimal steady-state~$x^*$ are equivalent~\cite{Grune_Muller_SCL16,Faulwasser_et_al_Aut17}.

To summarize the above, (strict) dissipativity with respect to the supply rate~\eqref{supply_rate_steady_state} and the set $\mathbb{X}^*=\{x^*\}$ both allows to conclude that the system is optimally operated at steady state and that the closed loop converges to $x^*$, i.e., the optimal operating regime is found.
An interesting question is to determine classes of systems that satisfy this dissipativity property. 
To this end, the followind results have been obtained in the literature.
It has first been noted in~\cite{Diehl_Amrit_Rawlings_TAC11} (compare also~\cite{Damm_et_al_preprint_Bayreuth12} for a rigorous proof) that linear systems with convex constraints (satisfying a Slater condition) and a strictly convex cost function~$\ell$ are strictly dissipative with respect to the supply rate~\eqref{supply_rate_steady_state} and the set $\mathbb{X}^*=\{x^*\}$, using a linear storage function\footnote{Nonnegativity of~$\lambda$ as required in Definition~\ref{def_dissipativity} then holds on any bounded set~$\mathbb{X}$.} $\lambda(x)=a^Tx+c$.
This result has been extended in~\cite{Damm_et_al_preprint_Bayreuth12,Grune_Guglielmi_17} to cost functions which are only convex (instead of strictly convex) in the state. 
In this case, the above strict dissipativity condition (with a quadratic storage function~$\lambda$) is equivalent to certain eigenvalue conditions on the system matrix. 
For linear systems with convex constraints and indefinite quadratic cost functions, the paper~\cite{Berberich_et_al_CSL18} develops conditions under which strict dissipativity with respect to the supply rate~\eqref{supply_rate_steady_state} and the set $\mathbb{X}^*=\{x^*\}$ is satisfied.
In this case, the optimal steady state is often located on the boundary of the constraint set.
Finally, in~\cite{Schmitt_et_al_ACC17}, it was shown that under some technical conditions, convex, (state-)monotone nonlinear systems with convex constraints and convex, (state-)monotone cost function are dissipative with respect to the supply rate~\eqref{supply_rate_steady_state}.

\section{Optimal periodic operation} \label{subsec_optimal_periodic_operation}
We now turn to the more general case of optimal periodic operation.
As described below, similar results as in Section~\ref{subsec_optimal_steady_state_operation} can be obtained.
To this end, two different types of dissipativity conditions have been developed in the literature, a periodic dissipativity condition and a dissipativity condition for a multi-step system.
Recently, it was shown that both these conditions are, in fact, equivalent to a standard dissipativity condition for system~\eqref{system}, analogous to Section~\ref{subsec_optimal_steady_state_operation}.
This will be discussed in detail in the following.

We first define the notion of a periodic orbit.
\begin{definition} \label{def_periodic_orbit}
A set of state/input pairs $\Pi = \{(x_0^p,u_0^p),\dots,(x_{P-1}^p,u_{P-1}^p)\}$ with $P\in\mathbb{N}$ is a \emph{feasible $P$-periodic orbit} of system~\eqref{system} if $(x_k^p,u_k^p)\in\mathbb{X}\times\mathbb{U}$ for all $k=0,\dots,P-1$, $x_{k+1}^p=f(x_k^p,u_k^p)$ for all $k=0,\dots,P-2$, and $x_0^p=f(x_{P-1}^p,u_{P-1}^p)$.  
\end{definition}
Given a periodic orbit $\Pi$, we denote its average cost by
\begin{align}
\ell_{\Pi}:=\frac{\sum_{k=0}^{P-1}\ell(x_k^p,u_k^p)}{P}. \label{av_cost_per_orbit} 
\end{align}
We can now define optimal periodic operation of a system as follows, analogous to optimal steady-state operation.
\begin{definition} \label{def_optimal_periodic_operation}
The system~(\ref{system}) is \emph{optimally operated at a periodic orbit $\Pi$}, if for each $x\in\mathbb{X}$ with $\mathcal{U}_{\infty}(x)\neq\emptyset$ and each $\mathbf{u}\in\mathcal{U}_{\infty}(x)$ the following holds:
\begin{align}
\liminf_{T\rightarrow\infty}\frac{\sum_{k=0}^T\ell(x_{\textbf{u}}(k;x),u(k))}{T+1} \geq \ell_{\Pi}. \label{condition_opt_periodic_op}
\end{align}
\end{definition}
Analogous to optimal steady-state operation, optimal periodic operation means that no other feasible input and state trajectory pair of system~\eqref{system} can result in an asymptotic average cost which is better than that of the periodic orbit~$\Pi$.
Clearly, if a system is optimally operated at some periodic orbit~$\Pi^*$, then $\Pi^*$ is the optimal periodic orbit for this system, i.e.,
\begin{align}
\ell_{\Pi^*} = \inf_{P\in\mathbb{N},\Pi\in S_{\Pi}^P}\ell_{\Pi}, \label{optimal_periodic_orbit}
\end{align}
where $S_{\Pi}^P$ denotes the set of all feasible $P$-periodic orbits.
Furthermore, note that the definition of optimal periodic operation contains the definition of optimal steady-state operation as a special case (for $P=1$).

As mentioned above, two different dissipativity conditions have separately been developed in the literature to analyse optimal periodic operation.
The first is a periodic dissipativity condition and has been proposed in\footnote{The first periodic dissipativity condition in the context of economic MPC has already been proposed in~\cite{Zanon_et_al_CDC13}.
However, besides the fact that the employed periodic storage functions are required to be linear as an extension of the strong duality condition in~\cite{Diehl_Amrit_Rawlings_TAC11}, the periodic dissipativity condition as formulated in~\cite{Zanon_et_al_CDC13} can only be satisfied by time-varying (periodic) systems, but not by time-invariant systems as in~\eqref{system}, compare~\cite[Remark~3.5]{Zanon_et_al_TAC17}.}~\cite{Zanon_et_al_TAC17}.

\begin{definition} \label{def_periodic_dissipativity}
The system~(\ref{system}) is $P$-periodically dissipative on $\mathbb{X}\times\mathbb{U}$ with respect to the supply rates $s_k(x,u)=\ell(x,u)-\ell(x_k^p,u_k^p)$, $k=0,\dots,P-1$, if there exist storage functions~$\lambda_k:\mathbb{R}^n\rightarrow\mathbb{R}_{\geq 0}$ for $k=0,\dots,P$ with $\lambda_P=\lambda_0$ such that the following inequality is satisfied for all $(x,u)\in\mathbb{X}\times\mathbb{U}$ with $f(x,u)\in\mathbb{X}$ and all $k=0,\dots,P-1$:
\begin{align}
\lambda_{k+1}(f(x,u))-\lambda_k(x)\leq s_k(x,u).  \label{dissipation_inequality_periodic}
\end{align}
If there exists $\rho\in\mathcal{K}_{\infty}$ such that~\eqref{dissipation_inequality_periodic} holds with $s_k(x,u)$ on the right hand side replaced by $s_k(x,u)-\rho(|(x,u)|_{\Pi})$, then system~(\ref{system}) is strictly dissipative with respect to the supply rates~$s_k$ and the periodic orbit~$\Pi$. 
\end{definition}

For $P=1$ this definition recovers the dissipativity condition of Section~\ref{subsec_optimal_steady_state_operation}, i.e., dissipativity of system~\eqref{system} with respect to the supply rate\footnote{\label{footnote_strict_periodic_diss}The definition of \emph{strict} $P$-periodic dissipativity with $P=1$ is slightly stronger than strict dissipativity with respect to the supply rate~\eqref{supply_rate_steady_state} and the set $\mathbb{X}^*=\{x^*\}$, since both strictness in state and input is required. Such a slightly stronger property is typically required in the context of optimal periodic operation in order to establish closed-loop convergence to the optimal periodic orbit, compare, e.g.,~\cite{Zanon_et_al_TAC17,Muller_Grune_Aut15}. Furthermore, in~\cite{Zanon_et_al_TAC17} a weaker variant of strict periodic dissipativity with $\rho(|(x,u)|_{\Pi})$ replaced by $\rho(|(x)|_{\Pi_{\mathbb{X}}})$ has been considered, where $\Pi_{\mathbb{X}}$ denotes the projection of $\Pi$ on $\mathbb{X}$, resulting in a slightly weaker closed-loop stability property.}~\eqref{supply_rate_steady_state}.
In~\cite{Zanon_et_al_TAC17}, it was shown that $P$-periodic dissipativity with respect to the supply rates $s_k(x,u)=\ell(x,u)-\ell(x_k^p,u_k^p)$ implies that the system~\eqref{system} is optimally operated at the periodic orbit~$\Pi$.
Analogous to the optimal steady-state case in Section~\ref{subsec_optimal_steady_state_operation}, strict $P$-periodic dissipativity with respect to the supply rates $s_k(x,u)=\ell(x,u)-\ell(x_k^p,u_k^p)$ and the periodic orbit~$\Pi$ implies a slightly stronger property than optimal periodic operation (uniform suboptimal operation off the periodic orbit~$\Pi$, compare~\cite{Zanon_et_al_TAC17,Muller_et_al_NMPC15} for a precise definition).
Furthermore, under the same strict periodic dissipativity condition, asymptotic stability of the optimal periodic orbit for the closed-loop system can be established if suitable periodic terminal constraints are added to problem~\eqref{economic_MPC_problem}, compare~\cite{Zanon_et_al_TAC17}.

The second dissipativity condition that has been used in the literature to examine the case of optimal periodic operation is based on the $P$-step system dynamics. 
Namely, define an extended state $\tilde{x}=(\tilde{x}_0,\dots,\tilde{x}_{P-1})\in\mathbb{X}^P$, input $\tilde{u}=(\tilde{u}_0,\dots,\tilde{u}_{P-1})\in\mathbb{U}^P$ and dynamics 
\begin{align}\tilde{x}(t+1) = \begin{bmatrix}
                      x_{\mathbf{\tilde{u}}}(1;\tilde{x}_{P-1}) \\
                      \dots \\
                      x_{\mathbf{\tilde{u}}}(P;\tilde{x}_{P-1})
                     \end{bmatrix}
               =\begin{bmatrix}
                      f(\tilde{x}_{P-1},\tilde{u}_0) \\
                      f(f(\tilde{x}_{P-1},\tilde{u}_0),\tilde{u}_1) \\
                      \dots 
                     \end{bmatrix}. \label{P-step_system}
\end{align}
Furthermore, define the cost function associated to the $P$-step system~\eqref{P-step_system} as $$\tilde{\ell}(\tilde{x},\tilde{u}):=\sum_{j=0}^{P-1}\ell(x_{\mathbf{\tilde{u}}}(j;\tilde{x}_{P-1}),\tilde{u}_j)$$ and $|(\tilde{x},\tilde{u})|_{\Pi}:=\sum_{j=0}^{P-1}|(x_{\mathbf{\tilde{u}}}(j;\tilde{x}_{P-1}),u_j)|_{\Pi}$.
In~\cite{Muller_et_al_NMPC15}, it was shown that dissipativity of the $P$-step system~\eqref{P-step_system} with respect to the supply rate 
\begin{align} 
s(\tilde{x},\tilde{u})=\tilde{\ell}(\tilde{x},\tilde{u})-P\ell_{\Pi} \label{supply_rate_P-step_system}
\end{align}
is sufficient and (under an additional controllability condition) also necessary for optimal periodic operation.
Similarly, strict dissipativity with respect to the supply rate~\eqref{supply_rate_P-step_system} and the set\footnote{Here, again the slightly stronger version of strict dissipativity compared to~\eqref{strict_dissipation_inequality} is needed, i.e., strictness in both state and input, compare Footnote~\ref{footnote_strict_periodic_diss}.} $\Pi$ is sufficient and (again under a suitable controllability condition) necessary for uniform suboptimal operation off the periodic orbit~$\Pi$.
Furthermore, the same strict dissipativity condition for the $P$-step system can be used to show closed-loop convergence to the optimal periodic orbit when applying economic MPC schemes with~\cite{Wabersich_et_al_ECC18} and without~\cite{Muller_Grune_Aut15} additional terminal constraints in~\eqref{economic_MPC_problem}. 
Interestingly, however, convergence guarantees cannot necessarily be given for a standard (one-step) MPC scheme but in general only when using a $P$-step MPC scheme, meaning that the first $P$ components of the optimal input sequence $u^\ast(\cdot;\hat{x})$ are applied, before problem~\eqref{economic_MPC_problem} is solved again (compare~\cite{Muller_Grune_Aut15}).

In a recent publication~\cite{Koehler_et_al_CSL18}, it was shown that under some technical conditions, the two above discussed different notions of (strict) dissipativity in the context of optimal periodic operation are equivalent.
In fact, they are equivalent to system~\eqref{system} being dissipative with respect to the supply rate
\begin{align} 
s(x,u)=\ell(x,u)-\ell_{\Pi},  \label{diss_periodic_one-step}
\end{align}
(respectively, strictly dissipative with respect to the supply rate~\eqref{diss_periodic_one-step} and the set~$\Pi$).
Note that this is a standard (one-step) dissipativity condition for system~\eqref{system}, as defined in Definition~\ref{def_dissipativity}.
The benefit of this result is that the cases of optimal steady-state and optimal periodic operation can now be treated within the same framework using standard dissipation inequalities (i.e., without having to define periodic dissipativity notions or multi-step system dynamics).
Also, the employed supply rates~\eqref{supply_rate_steady_state} and~\eqref{diss_periodic_one-step} are quite similar.
In particular, they are of the form $s(x,u)=\ell(x,u)-c$, where the constant $c$ is the asymptotic average cost of the optimal operating behavior (optimal steady-state or optimal periodic orbit).
As shown in the following section, such a dissipativity condition can also be used beyond optimal periodic operation.
Furthermore, when using the one-step dissipativity condition~\eqref{diss_periodic_one-step}, various of the previous assumptions (such as local controllability of the $P$-step system at the optimal periodic orbit) can be relaxed, and closed-loop convergence to the optimal periodic orbit can be established under certain conditions also for standard one-step MPC schemes without terminal constraints, compare~\cite{Koehler_et_al_CSL18}.

\section{General optimal operating conditions}  \label{subsec_general_optimal_operation}
The results of the previous two sections can be extended to optimal operating conditions which are more general than steady-state or periodic operation, as was recently shown in~\cite{Dong_Angeli_IJRNC18,Martin_et_al_ACC19}.
Namely, consider an arbitrary feasible state and input trajectory pair of system~\eqref{system}
\begin{align}
(x_0^*,u_0^*),(x_1^*,u_1^*),\dots \label{optimal_trajectory}
\end{align}
with corresponding (best) asymptotic average cost defined as
\begin{align}
\ell_{av} := \liminf_{T\rightarrow\infty}\frac{\sum_{k=0}^T\ell(x_k^*,u_k^*)}{T+1}.
\end{align}
Then, it was proven in~\cite{Dong_Angeli_IJRNC18} that if system~\eqref{system} is dissipative with respect to the supply rate
\begin{align} 
s(x,u)=\ell(x,u)-\ell_{av},  \label{diss_general_one-step}
\end{align}
the optimal asymptotic average performance is given by~$\ell_{av}$.
This means that for each $x\in\mathbb{X}$ with $\mathcal{U}_{\infty}(x)\neq\emptyset$ and each $\mathbf{u}\in\mathcal{U}_{\infty}(x)$ the following holds:
\begin{align*}
\liminf_{T\rightarrow\infty}\frac{\sum_{k=0}^T\ell(x_{\textbf{u}}(k;x),u(k))}{T+1} \geq \ell_{av}.
\end{align*}
The proof of this fact is analogous to the one provided in Section~\ref{subsec_optimal_steady_state_operation} for optimal steady-state operation, replacing $\ell(x^*,u^*)$ by~$\ell_{av}$.
It is easy to see that both optimal steady-state and optimal periodic operation are included as a special case, when using a constant or periodic trajectory in~\eqref{optimal_trajectory}, respectively.

Furthermore, a strict version of this dissipativity condition can again be used in order to show closed-loop convergence to the optimal regime of operation.
Namely, denote by $\Pi$ the closure of the set of all points of the trajectory~\eqref{optimal_trajectory}, i.e., $\Pi:=\textnormal{cl}\{(x_0^*,u_0^*),(x_1^*,u_1^*),\dots\}$, and the projection of $\Pi$ on $\mathbb{X}$ by $\Pi_{\mathbb{X}}$, i.e., $\Pi_{\mathbb{X}}:=\textnormal{cl}\{x_0^*,x_1^*,\dots\}$. 
Then, if a suitable terminal region and terminal cost function is employed and some technical assumptions hold, strict dissipativity of system~\eqref{system} with respect to the supply rate~\eqref{diss_general_one-step} and the set~$\Pi_{\mathbb{X}}$ implies asymptotic stability of the set~$\Pi_{\mathbb{X}}$ for the resulting closed-loop system.

A different dissipativity condition has recently been proposed in~\cite{Martin_et_al_ACC19} in the context of optimal set operation, using parametric storage functions. 
Namely, consider some control invariant set $\overline{\mathbb{X}}\subseteq\mathbb{X}$ and define $\overline{\mathbb{Z}}:=\{(x,u)\in\overline{\mathbb{X}}\times\mathbb{U}:f(x,u)\in\overline{\mathbb{X}}\}$ as well as $\overline{\mathcal{U}}_{\infty}(\bar{x}):=\{\textbf{u}\in\mathbb{U}^N:x_{\textbf{u}}(k;x)\in\overline{\mathbb{X}}~\textnormal{for all}~k\geq 0\}$.
Now consider the dissipation inequality
\begin{align}
\lambda_{f(\bar{x},\bar{u})}(f(x,u))-\lambda_{\bar{x}}(x) \leq \ell(x,u)-\ell(\bar{x},\bar{u}), \label{parametric_diss_inequality}
\end{align}
where $(\bar{x},\bar{u})\in\overline{\mathbb{Z}}$. 
As shown in~\cite{Martin_et_al_ACC19}, if~\eqref{parametric_diss_inequality} is satisfied for all $(x,u)\in\mathbb{X}\times\mathbb{U}$ with $f(x,u)\in\mathbb{X}$ and all $(\bar{x},\bar{u})\in\overline{\mathbb{Z}}$, then system~\eqref{system} is optimally operated at the set~$\overline{\mathbb{Z}}$.
This means that for each $x\in\mathbb{X}$ with $\mathcal{U}_{\infty}(x)\neq\emptyset$, each $\mathbf{u}\in\mathcal{U}_{\infty}(x)$, each $\bar{x}\in\overline{\mathbb{X}}$ with $\overline{\mathcal{U}}_{\infty}(\bar{x})\neq\emptyset$, and each $\bar{\mathbf{u}}\in\overline{\mathcal{U}}_{\infty}(\bar{x})$ the following holds:
\begin{align*}
\liminf_{T\rightarrow\infty}\frac{\sum_{k=0}^T\ell(x_{\textbf{u}}(k;x),u(k))}{T+1} \geq \liminf_{T\rightarrow\infty}\frac{\sum_{k=0}^T\ell(\bar{x}_{\bar{\mathbf{u}}}(k;\bar{x}),\bar{u}(k))}{T+1}, 
\end{align*}
i.e., the optimal asymptotic average performance is given by the asymptotic average performance of an arbitrary trajectory inside the set~$\overline{\mathbb{Z}}$.
Again, also the converse statement is true under an additional controllability assumption~\cite{Martin_et_al_ACC19}.

The dissipativity condition~\eqref{parametric_diss_inequality} uses a storage function which is parametrized by states~$\bar{x}\in\overline{\mathbb{X}}$ and can be seen as a generalization of the periodic dissipativity condition in Definition~\ref{def_periodic_dissipativity}.
Namely, while Definition~\ref{def_periodic_dissipativity} uses $P$ storage functions (i.e., parameterized by the optimal periodic orbit in terms of its period length), the storage function in~\eqref{parametric_diss_inequality} is parameterized along arbitrary trajectories in the set~$\overline{\mathbb{X}}$. 
Using suitable terminal equality constraints specified by a trajectory in~$\overline{\mathbb{X}}$, it can be shown that the dissipativity conditon~\eqref{parametric_diss_inequality} strengthened to strict dissipativity with respect to the set~$\overline{\mathbb{X}}$ ensures closed-loop convergence to~$\overline{\mathbb{X}}$ as desired~\cite{Martin_et_al_ACC19}.

\section{Computation of storage functions} \label{subsec_comp_storage_fct}
In this section, we briefly discuss how the above dissipativity properties can be verified, i.e., how suitable storage functions satisfying the relevant dissipation inequalities can be computed.

First, we note that also without explicitly verifying dissipativity, the above results can be used as an a posteriori guarantee that - given a suitably designed economic MPC scheme - the closed-loop system "`does the right thing"', i.e., converges to the optimal operating behavior: (strict) optimal steady-state/periodic/set operation implies (strict) dissipativity, which in turn can be used to conclude closed-loop convergence.

On the other hand, in order to obtain a priori guarantees about the optimal operating conditions and the closed-loop behavior, (strict) dissipativity needs to be verified.
To this end, different approaches are available for different system classes (although it has to be mentioned that no systematic procedure is available for general nonlinear systems).
As has already been discussed at the end of Section~\ref{subsec_optimal_steady_state_operation}, for linear systems linear or quadratic storage functions can be computed under certain conditions.
If nonlinear polynomial systems subject to polynomial cost and constraints are considered, sum-of-squares (SOS) programming can be employed to verify dissipativity, compare, e.g.,~\cite{EbenbauerAllgower06}. 
In the context of economic MPC, this method has, e.g., been applied in~\cite{Faulwasser_et_al_CDC14,Dong_Angeli_IJRNC18} for verifying dissipativity in case of optimal steady-state and optimal periodic operation, respectively.
In these examples, the optimal steady-state cost $\ell(x^*,u^*)$ and cost of the optimal periodic orbit $\ell_{\Pi}$, appearing in the supply rates~\eqref{supply_rate_steady_state} and~\eqref{diss_periodic_one-step}, respectively, had been assumed \emph{known} (i.e., precomputed before verifying dissipativity).

On the other hand, in general the optimal operating behavior might be \emph{unknown} a priori and hence the optimal average cost appearing in the supply rates~\eqref{supply_rate_steady_state}, \eqref{diss_periodic_one-step}, and \eqref{diss_general_one-step} has to be computed together with a suitable storage function.
To this end, a computational procedure has recently been proposed concurrently in~\cite{Berberich_et_al_Aut19} and~\cite{PirkelmannAngeliGrune19}.
Namely, there the following optimization problem has been considered:
\begin{align}
&\textnormal{maximize}_{c\in\mathbb{R},\lambda\in\Lambda}~~~c \label{opt_prob_determine_lambda}\\ 
&\textnormal{s.t.}~~\ell(x,u)-c+\lambda(x)-\lambda(f(x,u)) \geq 0~~\textnormal{for all}~(x,u)\in\mathbb{Z}, \label{constraint_dissipativity}
\end{align}
where $\mathbb{Z}:=\{(x,u)\in\mathbb{X}\times\mathbb{U}:f(x,u)\in\mathbb{X}\}$ and $\Lambda\subseteq C(\mathbb{R}^n)$ is a given set of functions.
Constraint~\eqref{constraint_dissipativity} ensures that system~\eqref{system} is dissipative with respect to the supply rate
\begin{align}
s(x,u)=\ell(x,u)-c. \label{supply_rate_comparison}
\end{align}
Denote by $c^*$ and $\lambda^*$ the optimizers to this problem (assuming they exist).
Hence, using the same arguments as above, it follows that $c^*$ is a lower bound on the best achievable asymptotic average cost, compare~\cite{Dong_Angeli_IJRNC18}.
Unfortunately, problem~\eqref{opt_prob_determine_lambda}--\eqref{constraint_dissipativity} can, in general, not be solved efficiently, since (even for a fixed, finite parameterization of~$\Lambda$) it is a semi-infitite optimization problem.
Nevertheless, for polynomial system dynamics $f$ and polynomial cost function~$\ell$, sum-of-squares programming can again be employed to efficiently solve a relaxed formulation of problem~\eqref{opt_prob_determine_lambda}--\eqref{constraint_dissipativity}, compare~\cite{Berberich_et_al_Aut19,PirkelmannAngeliGrune19}.
This is straightforward in case of no constraints, i.e., $\mathbb{X}=\mathbb{R}^n$ and $\mathbb{U}=\mathbb{R}^m$.
In case that state and/or input constraints are present which are given in terms of polynomial inequalities, the Positivstellensatz or S-procedure can be used in order to obtain again a relaxed formulation of problem~\eqref{opt_prob_determine_lambda}--\eqref{constraint_dissipativity}, which can efficiently be solved by SOS techniques~\cite{Berberich_et_al_Aut19,PirkelmannAngeliGrune19}.
These methods can also be used to verify approximate dissipativity for nonpolynomial systems by considering (polynomial) Taylor approximations~\cite{PirkelmannAngeliGrune19}.

In~\cite{Berberich_et_al_Aut19}, it was additionally considered how \emph{strict} dissipativity can be verified using problem~\eqref{opt_prob_determine_lambda}--\eqref{constraint_dissipativity}.
Namely, consider the set of points for which the constraint~\eqref{constraint_dissipativity} is satisfied with equality for $c=c^*$ and $\lambda=\lambda^*$, i.e.,
\begin{align}
M:=\{(x,u)\in\mathbb{Z}:\ell(x,u)-c^*+\lambda^*(x)-\lambda^*(f(x,u))=0\}. \label{set_M_strict_diss}
\end{align}
The following results can now be obtained.
If $M$ is a periodic orbit, then $\ell_M=c^*$ and system~\eqref{system} is strictly dissipative with respect to the supply rate~\eqref{supply_rate_comparison} and the periodic orbit~$M$.
Conversely, if system~\eqref{system} is strictly dissipative with respect to the supply rate~\eqref{diss_periodic_one-step} and some periodic orbit~$\Pi$ with storage function $\lambda\in\Lambda$, then $c^*=\ell_{\Pi}$ and $\Pi\subseteq M$.
Extensions to more general operating conditions are also possible.
Furthermore, in case that the above discussed relaxations via SOS programming are employed, similar sufficient conditions for strict dissipativity can be obtained (based on a set defined similarly to~$M$ in~\eqref{set_M_strict_diss} and an additional complementary slackness condition), albeit necessity does no longer hold, compare~\cite{Berberich_et_al_Aut19}.

\section{Time-varying case} \label{subsec_time_varying}
In this section, we discuss how the above results can be extended to the time-varying case, i.e., time-varying system dynamics of the form
\begin{align}
x(k+1)=f(k,k(k),u(k)), \quad x(t_0)=x_0, \label{time-varying-system}
\end{align}
together with a time-varying cost function $\ell(k,x(k),u(k))$ and constraint sets $\mathbb{X}(k)$ and $\mathbb{U}(k)$.
While the main insights and proof techniques carry over, various technical subtleties become more involved in this case.

The first issue is how to properly define the optimal operating behavior in the time-varying case.
To this end, two different approaches have been used in the literature.
In~\cite{Alessandretti_et_al_Aut16}, an averaged criterion similar to the time-invariant case has been used.
Namely, the system is optimally operated at some feasible state and input trajectory pair $(x^*(\cdot),u^*(\cdot))$, if
for all $x\in\mathbb{X}(k)$ with\footnote{Analogous to the time-invariant case, $\mathcal{U}_{\infty}(k,x)$ denotes the (time-varying) set of all feasible input sequences of infinite length for a given initial state $x$ at time~$k$.} $\mathcal{U}_{\infty}(k,x)\neq\emptyset$ and each $\mathbf{u}\in\mathcal{U}_{\infty}(k,x)$ the following holds:
\begin{align}
\liminf_{T\rightarrow\infty}\frac{\sum_{k=0}^T\ell(k,x_{\textbf{u}}(k;x),u(k))-\ell(k,x^*(k),u^*(k))}{T+1} \geq 0. \label{condition_opt_time-varying_operation}
\end{align}
Similar to the previous sections, this means that no feasible state and input trajectory pair results in a better asymptotic average cost than the pair $(x^*(\cdot),u^*(\cdot))$.
Alternatively, a slightly stronger definition of optimal system operation is employed in~\cite{Grune_et_al_time-varying-dissipativity-turnpike_18,GrunePirkelmann_OCAM19}, where the inequality~\eqref{condition_opt_time-varying_operation} is evoked in a non-averaged sense.
In particular, optimal system operation at some feasible state and input trajectory pair $(x^*(\cdot),u^*(\cdot))$ is given if for all $x\in\mathbb{X}(k)$ with $\mathcal{U}_{\infty}(k,x)\neq\emptyset$ and each $\mathbf{u}\in\mathcal{U}_{\infty}(k,x)$ the following holds:
\begin{align}
\liminf_{T\rightarrow\infty}\sum_{k=0}^T\ell(k,x_{\textbf{u}}(k;x),u(k))-\ell(k,x^*(k),u^*(k)) \geq 0. \label{condition_opt_time-varying_operation_overtaking_optimality} 
\end{align}
This definition is related to the concept of \emph{overtaking optimality} in the sense that the (cumulative) cost encountered along the optimal trajectory pair $(x^*(\cdot),u^*(\cdot))$ is "`overtaken"' by the (cumulative) cost encountered along any other feasible state and input trajectory pair at some point in time.
 
Again, a suitable (time-varying) dissipativity condition can be employed to classify optimal system operation and to analyze the closed-loop behavior of time-varying economic MPC schemes.
\begin{definition} \label{def_dissipativity_time_varying}
The system~\eqref{time-varying-system} is dissipative with respect to the supply rate $s:\mathbb{N}_0\times\mathbb{R}^n\times\mathbb{R}^m\rightarrow\mathbb{R}$ if there exists a storage function~$\lambda:\mathbb{N}_0\times\mathbb{R}^n\rightarrow\mathbb{R}_{\geq 0}$ such that the following inequality is satisfied for all $k\in\mathbb{N}_0$ and all $(x,u)\in\mathbb{X}(k)\times\mathbb{U}(k)$ with $f(k,x,u)\in\mathbb{X}(k+1)$:
\begin{align}
\lambda(k+1,f(k,x,u))-\lambda(k,x)\leq s(k,x,u).  \label{dissipation_inequality_time-varying}
\end{align}
If there exists $\rho\in\mathcal{K}_{\infty}$ such that~\eqref{dissipation_inequality_time-varying} holds with $s(k,x,u)$ on the right hand side replaced by $s(k,x,u)-\rho(|(x,u)|_{(x^*(k),u^*(k))})$, then system~(\ref{time-varying-system}) is strictly dissipative with respect to the supply rates~$s$ and the trajectory pair $(x^*(\cdot),u^*(\cdot))$.  
\end{definition}

Using the same arguments as in the time-invariant case, one can show that dissipativity with respect to the supply rate $s(k,x,u)=\ell(k,x,u)-\ell(k,x^*(k),u^*(k))$ is a sufficient condition for optimal system operation at the trajectory pair $(x^*(\cdot),u^*(\cdot))$ in the sense of~\eqref{condition_opt_time-varying_operation}, compare~\cite{Alessandretti_et_al_Aut16}.
However, this is not necessarily the case when using the stronger, non-averaged version~\eqref{condition_opt_time-varying_operation_overtaking_optimality} of optimal system operation.
Furthermore, \emph{strict} dissipativity with respect to the supply rate $s(k,x,u)=\ell(k,x,u)-\ell(k,x^*(k),u^*(k))$ and the the trajectory pair $(x^*(\cdot),u^*(\cdot))$ can (together with some technical assumptions) be employed to establish (i) time-varying turnpike properties~\cite{Grune_et_al_time-varying-dissipativity-turnpike_18} and (ii) closed-loop (practical) convergence to the optimal trajectory pair $(x^*(\cdot),u^*(\cdot))$, both for economic MPC schemes with\footnote{In~\cite{Alessandretti_et_al_Aut16}, the additional condition that $\lambda$ is constant along $x^*(\cdot)$ is imposed. Note that from~\eqref{dissipation_inequality_time-varying} together with the definition of the supply rate $s(k,x,u)=\ell(k,x,u)-\ell(k,x^*(k),u^*(k))$, it follows that $\lambda$ converges to a constant value~$\bar\lambda$ when evaluated along $x^*(\cdot)$, i.e., $\lim_{k\rightarrow\infty}\lambda(k,x^*(k))=\bar\lambda$, but not necessarily that $\lambda$ is constant along $x^*(\cdot)$ for all $k$.}~\cite{Alessandretti_et_al_Aut16} and without~\cite{GrunePirkelmann_OCAM19} suitable time-varying terminal cost and terminal region.

\section{Conclusions}   \label{subsec_conclusions}
Dissipativiy has turned out to play a crucial role in the context of economic model predictive control, both for determining the optimal operating behavior of a system as well as for the analysis of the closed-loop dynamics.
As shown above, to this end a dissipativity condition with supply rate of the form
\begin{align}
s(x,u)=\ell(x,u)-c \label{supply_rate_conclusions}
\end{align}
can be used\footnote{As discussed in the previous sections, this holds for the time-invariant case. The supply rates in the time-varying case or when using parameterized storage functions (compare Sections~\ref{subsec_time_varying} and~\ref{subsec_general_optimal_operation}, respectively), are of a similar structure and hence allow for the same interpretation as discussed in the following for supply rate~\eqref{supply_rate_conclusions}.}, where the constant $c$ is the asymptotic average cost of the optimal operating behavior.
This holds true for the basic case of optimal steady-state operation, for optimal periodic operation, and also for more general optimal regimes of operation.
This allows for the following high-level intuition/interpretation.
Recalling the "`energy"' interpretation of dissipativity given in Section~\ref{subsec_dissipativity}, a negative value of the supply rate~$s$ corresponds to "`extracting energy"' from the system. 
In case of the supply rate~\eqref{supply_rate_conclusions}, this is the case for all points $(x,u)$ which have a lower cost than the optimal asymptotic average cost~$c$. 
Since dissipativity means that we cannot extract an infinite amount of energy from the system, on average (asymptotically) the supply rate must be nonnegative, which by~\eqref{supply_rate_conclusions} means that the optimal asymptotic average cost cannot be lower than~$c$.

Recently, some extensions and generalizations of the above results have been proposed, e.g., in the context of uncertain systems~\cite{Bayer_et_al_Aut18}, for discounted optimal control problems~\cite{Grune_et_al_NOLCOS16,Muller_Grune_CDC17}, and in distributed~\cite{KohlerP_et_al_ECC19} and multi-objective~\cite{Stieler_PhD_thesis} settings. 
Here, however, the picture of the interplay of suitable dissipativity conditions, optimal system operation and economic MPC schemes is still much less complete and allows for many interesting directions for future research.

\bibliographystyle{plain}
\bibliography{biblio_Muller}

\begin{thebibliography}{10}

\bibitem{Alessandretti_et_al_Aut16}
A.~Alessandretti, A.~P. Aguiar, and C.~N. Jones.
\newblock On convergence and performance certification of a continuous-time
  economic model predictive control scheme with time-varying performance index.
\newblock {\em Automatica}, 68:305 -- 313, 2016.

\bibitem{Amrit_Rawlings_Angeli_AnnRevCont_11}
R.~Amrit, J.~B. Rawlings, and D.~Angeli.
\newblock Economic optimization using model predictive control with a terminal
  cost.
\newblock {\em Annual Reviews in Control}, 35(2):178--186, 2011.

\bibitem{Angeli_Amrit_Rawlings_TAC12}
D.~Angeli, R.~Amrit, and J.~B. Rawlings.
\newblock On average performance and stability of economic model predictive
  control.
\newblock {\em IEEE Transactions on Automatic Control}, 57(7):1615--1626, 2012.

\bibitem{Bayer_et_al_Aut18}
F.~A. Bayer, M.~A. M\"uller, and F.~Allg\"ower.
\newblock On optimal system operation in robust economic {MPC}.
\newblock {\em Automatica}, 88:98 -- 106, 2018.

\bibitem{Berberich_et_al_CSL18}
J.~Berberich, J.~K\"ohler, F.~Allg\"ower, and M.~A. M\"uller.
\newblock Indefinite linear quadratic optimal control: Strict dissipativity and
  turnpike properties.
\newblock {\em IEEE Control Systems Letters}, 2(3):399--404, July 2018.

\bibitem{Berberich_et_al_Aut19}
J.~Berberich, J.~K\"ohler, F.~All\"ower, and M.~A. M\"uller.
\newblock Dissipativity properties in constrained optimal control: a
  computational approach.
\newblock {\em Automatica}, 2018.
\newblock Submitted. Available online:
  \url{https://www.ist.uni-stuttgart.de/de/institut/team/PDFs_MA-Seiten/JB/Dissipativity_Automatica.pdf}.

\bibitem{Byrnes_Lin_TAC94}
C.~I. Byrnes and W.~Lin.
\newblock Losslessness, feedback equivalence, and the global stabilization of
  discrete-time nonlinear systems.
\newblock {\em IEEE Transactions on Automatic Control}, 39(1):83--98, 1994.

\bibitem{Damm_et_al_preprint_Bayreuth12}
T.~Damm, L.~Gr\"une, M.~Stieler, and K.~Worthmann.
\newblock An exponential turnpike theorem for dissipative discrete time optimal
  control problems.
\newblock {\em SIAM Journal on Control and Optimization}, 52(3):1935--1957,
  2014.

\bibitem{Diehl_Amrit_Rawlings_TAC11}
M.~Diehl, R.~Amrit, and J.~B. Rawlings.
\newblock A {Lyapunov} function for economic optimizing model predictive
  control.
\newblock {\em IEEE Transactions on Automatic Control}, 56(3):703--707, 2011.

\bibitem{Dong_Angeli_IJRNC18}
Z.~Dong and D.~Angeli.
\newblock Analysis of economic model predictive control with terminal penalty
  functions on generalized optimal regimes of operation.
\newblock {\em International Journal of Robust and Nonlinear Control},
  28(16):4790--4815, 2018.

\bibitem{EbenbauerAllgower06}
C.~Ebenbauer and F.~Allg\"ower.
\newblock Analysis and design of polynomial control systems using dissipation
  inequalities and sum of squares.
\newblock {\em Computers \& Chemical Engineering}, 30(10):1590 -- 1602, 2006.
\newblock Papers form Chemical Process Control VII.

\bibitem{Ellis17a}
M.~Ellis, M.~Liu, and P.~Christofides.
\newblock {\em Economic Model Predictive Control: Theory, Formulations and
  Chemical Process Applications}.
\newblock Springer, Berlin, 2017.

\bibitem{Faulwasser_Grune_Muller_18}
T.~Faulwasser, L.~Gr\"une, and M.~A. M\"uller.
\newblock Economic nonlinear model predictive control.
\newblock {\em Foundations and Trends\textsuperscript{\textregistered} in
  Systems and Control}, 5(1):1--98, 2018.

\bibitem{Faulwasser_et_al_CDC14}
T.~Faulwasser, M.~Korda, C.~N. Jones, and D.~Bonvin.
\newblock Turnpike and dissipativity properties in dynamic real-time
  optimization and economic {MPC}.
\newblock In {\em Proceedings of the 53rd IEEE Conference on Decision and
  Control}, pages 2734--2739, 2014.

\bibitem{Faulwasser_et_al_Aut17}
T.~Faulwasser, M.~Korda, C.~N. Jones, and D.~Bonvin.
\newblock On turnpike and dissipativity properties of continuous-time optimal
  control problems.
\newblock {\em Automatica}, 81:297 -- 304, 2017.

\bibitem{Gruene_econ_MPC_sub_11}
L.~Gr\"une.
\newblock Economic receding horizon control without terminal constraints.
\newblock {\em Automatica}, 49(3):725--734, 2013.

\bibitem{Grune_Guglielmi_17}
L.~Gr\"une and R.~Guglielmi.
\newblock Turnpike properties and strict dissipativity for discrete time linear
  quadratic optimal control problems.
\newblock {\em SIAM Journal on Control and Optimization}, 56(2):1282--1302,
  2018.

\bibitem{Grune_et_al_NOLCOS16}
L.~Gr\"une, C.~M. Kellett, and S.~R. Weller.
\newblock On a discounted notion of strict dissipativity.
\newblock In {\em Proceedings of the 10th IFAC Symposium on Nonlinear Control
  Systems}, pages 247--252, 2016.

\bibitem{Grune_Muller_SCL16}
L.~Gr\"une and M.~A. M\"uller.
\newblock On the relation between strict dissipativity and turnpike properties.
\newblock {\em System \& Control Letters}, 90:45--53, 2016.

\bibitem{GrunePirkelmann_OCAM19}
L.~Gr\"une and S.~Pirkelmann.
\newblock Economic model predictive control for time-varying system:
  Performance and stability results.
\newblock {\em Optimal Control Applications and Methods}, 2019.

\bibitem{Grune_et_al_time-varying-dissipativity-turnpike_18}
L.~Gr\"une, S.~Pirkelmann, and M.~Stieler.
\newblock Strict dissipativity implies turnpike behavior for time-varying
  discrete time optimal control problems.
\newblock In {\em Control Systems and Mathematical Methods in Economics: Essays
  in Honor of Vladimir M. Veliov}, volume 687 of {\em Lecture Notes in
  Economics and Mathematical Systems}, pages 195--218. Springer, Cham, 2018.

\bibitem{Grune_Stieler_14}
L.~Gr\"une and M.~Stieler.
\newblock Asymptotic stability and transient optimality of economic {MPC}
  without terminal constraints.
\newblock {\em Journal of Process Control}, 24(8):1187--1196, 2014.

\bibitem{Koehler_et_al_CSL18}
J.~K\"ohler, M.~A. M\"uller, and F.~Allg\"ower.
\newblock On periodic dissipativity notions in economic model predictive
  control.
\newblock {\em IEEE Control Systems Letters}, 2(3):501--506, 2018.

\bibitem{KohlerP_et_al_ECC19}
P.~N. K\"ohler, M.~A. M\"uller, and F.~Allg\"ower.
\newblock Approximate dissipativity and performance bounds for interconnected
  systems.
\newblock In {\em Proceedings of the European Control Conference (ECC)}, pages
  787--792, 2019.

\bibitem{Martin_et_al_ACC19}
T.~Martin, P.~N. K\"ohler, and F.~Allg\"ower.
\newblock Dissipativity and economic model predictive control for optimal set
  operation.
\newblock In {\em Proceedings of the American Control Conference (ACC)}, 2019.

\bibitem{Muller_Angeli_Allgower_TAC13}
M.~A. M\"uller, D.~Angeli, and F.~Allg\"ower.
\newblock On necessity and robustness of dissipativity in economic model
  predictive control.
\newblock {\em IEEE Transactions on Automatic Control}, 60(6):1671--1676, 2015.

\bibitem{Muller_Grune_Aut15}
M.~A. M\"uller and L.~Gr\"une.
\newblock Economic model predictive control without terminal constraints for
  optimal periodic behavior.
\newblock {\em Automatica}, 70:128--139, 2016.

\bibitem{Muller_Grune_CDC17}
M.~A. M\"uller and L.~Gr\"une.
\newblock On the relation between dissipativity and discounted dissipativity.
\newblock In {\em Proc. of the 56th IEEE Conference on Decision and Control
  (CDC)}, pages 5570--5575, 2017.

\bibitem{Muller_et_al_NMPC15}
M.~A. M\"uller, L.~Gr\"une, and F.~Allg\"ower.
\newblock On the role of dissipativity in economic model predictive control.
\newblock In {\em Proceedings of the 5th IFAC Conference on Nonlinear Model
  Predictive Control}, pages 110--116, 2015.

\bibitem{PirkelmannAngeliGrune19}
S.~Pirkelmann, D.~Angeli, and L.~Gr\"une.
\newblock Approximate computation of storage functions for discrete-time
  systems using sum-of-squares techniques.
\newblock In {\em Preprint, University of Bayreuth}, 2019.

\bibitem{Schmitt_et_al_ACC17}
M.~{Schmitt}, C.~{Ramesh}, P.~{Goulart}, and J.~{Lygeros}.
\newblock Convex, monotone systems are optimally operated at steady-state.
\newblock In {\em 2017 American Control Conference (ACC)}, pages 2662--2667,
  May 2017.

\bibitem{Stieler_PhD_thesis}
M.~Stieler.
\newblock {\em Performance Estimates for Scalar and Multiobjective Model
  Predictive Control Schemes}.
\newblock PhD thesis, University of Bayreuth, 2018.

\bibitem{Wabersich_et_al_ECC18}
K.P. Wabersich, F.A. Bayer, M.~A. M\"uller, and F.~Allg\"ower.
\newblock Economic model predictive control for robust periodic operation with
  guaranteed closed-loop performance.
\newblock In {\em Proceedings of the European Control Conference}, pages
  507--513, 06 2018.

\bibitem{Willems_Dissipativity_PartI_72}
J.~C. Willems.
\newblock Dissipative dynamical systems - part i: {G}eneral theory.
\newblock {\em Archive for Rational Mechanics and Analysis}, 45(5):321--351,
  1972.

\bibitem{Zanon_et_al_CDC13}
M.~Zanon, S.~Gros, and M.~Diehl.
\newblock A {L}yapunov function for periodic economic optimizing model
  predictive control.
\newblock In {\em Proceedings of the 52nd IEEE Conference on Decision and
  Control}, pages 5107--5112, 2013.

\bibitem{Zanon_et_al_TAC17}
M.~Zanon, L.~Gr\"une, and M.~Diehl.
\newblock Periodic optimal control, dissipativity and {MPC}.
\newblock {\em IEEE Transactions on Automatic Control}, 62(6):2943--2949, 2017.

\end{thebibliography}

\end{document}